\documentclass[11pt]{article}
\usepackage[mathscr]{eucal}
\usepackage{theorem,amsmath,amssymb}
\usepackage{hyperref,graphicx}

\usepackage[textwidth=17.5truecm,textheight=24.0truecm]{geometry}

\begin{document}


{~}\vskip -0.cm \noindent
\begin{center}
{\LARGE A note on ``Algebraic approach to Casimir force between two\vspace{0.07cm}\\
$\delta$-like potentials'' (K. Ziemian, Ann.\! Henri Poincar\'e, Online First, 2021)}
\end{center}
\vspace{0.55truecm}
\begin{center}
Davide Fermi \\
\vspace{0.1truecm}
Dipartimento di Matematica, Universit\`a di Roma ``La Sapienza'',\\
Piazzale Aldo Moro 5, I-00185 Roma, Italy\\
e--mail address: fermidavide@gmail.com\\
\vspace{0.5truecm}
Livio Pizzocchero \\
\vspace{0.1truecm}
Dipartimento di Matematica, Universit\`a di Milano\\
Via C. Saldini 50, I-20133 Milano, Italy\\
and Istituto Nazionale di Fisica Nucleare, Sezione di Milano, Italy \\
e--mail address: livio.pizzocchero@unimi.it \vspace{0.7truecm}\\
\end{center}
\begin{abstract}
We comment on the recent work \cite{Z}, and on its relations with our
papers \cite{FP,F} cited therein. In particular we show that, contrarily to what
stated in \cite{Z}, the Casimir energy density determined therein in the case of
a single delta-like singularity coincides with the energy density
obtained previously in our paper \cite{FP} using a different approach.
\end{abstract}
\vspace{0.65truecm}

\section{Introduction}
In his recent paper \cite{Z} K. Ziemian considers a quantized scalar field
in presence of an external non-local pseudo-potential, modeling two extended impurities.
When a certain scaling parameter is sent to zero, the impurities
become point-like and the pseudo-potential turns into
a local potential consisting of two separate delta interactions; this
construction is similar to that described
in the book \cite{A} by Albeverio \textsl{et al.},
indicated in \cite{Z} as a basic reference. \par
Both for extended and for point-like impurities, Ziemian determines
in \cite{Z} the vacuum expectation value (VEV) of
the total energy and of the energy density for the
quantized scalar field, using Herdegen's approach \cite{H,HS} to Casimir physics
(the physics of vacuum states). Notice that the scalar field theory considered
in these references
has conformal parameter $\xi=0$ (as for the meaning of $\xi$ see, {\sl e.g.}, the book
by Bytsenko \textsl{et al.} \cite{M} or our book \cite{FPB}, together with the references cited therein).
Again in \cite{Z}, the limit case of a single impurity is treated sending
to infinity the position of the other impurity.
For a single point-like impurity, the VEV of the total energy
is found to be divergent (see Section 7 of \cite{Z})
while a finite, explicit expression is obtained for the VEV of the
energy density (see Section 8 of \cite{Z}, especially Eq. (106) therein).\par
The Casimir physics of scalar fields in presence
of point-like impurities was investigated in a number
of previous works, mentioned in Section 9 of
\cite{Z}. In particular, the case of a quantized
scalar field with a single point-like impurity was
discussed in our papers \cite{FP,F},
making reference to the zeta regularization
approach to Casimir physics (see again the books
\cite{M,FPB} and the references cited therein, accounting
for the long story of this approach). In \cite{FP}
we determined the VEV of all components of the stress-energy
tensor and,
in particular, of the energy density
(for arbitrary values of the conformal parameter $\xi$).
In \cite{F}, one of us
(D.F.) examined the VEV of the total energy
for the same system.
\vfill \eject \noindent
{~}
\vskip 0.4cm \noindent
Concerning our works \cite{FP,F}, Ziemian
makes  the following statements in Section
9 of \cite{Z} (note the use of quotation marks for literal citations):\vspace{0.1cm}
\begin{itemize}
\item[(I)] In the case of a single point-like
impurity, Ziemian claims that his formula
for the VEV of the energy density (namely, Eq. (106) of \cite{Z})
is ``very different'' from the corresponding formula
of our work \cite{FP} ({\sl viz.}, Eq. (60) of our paper with $\xi=0$).
As a comment on this alleged discrepancy, Ziemian
states that he could not find a reason
``why this circuitous procedure [i.e., the method of our paper \cite{FP}]
should yield back the correct local
expression for energy''.\vspace{0.1cm}
\item[(II)] Again in the case of a single point-like
impurity, Zieman mentions that one of us attempted
in \cite{F} to cure the divergence of the total
energy VEV using an approach that, in Ziemian's words,
``does not solve the problem''.\vspace{0.1cm}
\end{itemize}
The aims of this note are: to point out two minor computational errors,
that we have detected in \cite{Z}; to disprove Ziemian's claim in (I) showing
that, \textsl{if amended from the above
minor errors, the expression of \cite{Z}
for the energy density VEV with a single point-like impurity
coincides with the expression previously given in our paper
\cite{FP}};
to put Ziemian's statement (II) under a perspective that we believe
to be more appropriate. All these issues are discussed in Section \ref{disc} (and
in the Appendix);
let us mention that, in spite of the problems indicated in this note,
we believe \cite{Z} to be an interesting contribution to investigations
on Casimir physics.
\vspace{0.4cm}
\section{A discussion on \cite{Z}}
\label{disc}
Our remarks on Ziemian's paper are contained in the forthcoming items (A)-(D).
In particular, (C) and (D) are, respectively, our rebuttals to Ziemian's statements
(I) and (II) (reported in the Introduction of this note).
Some additional details can be found in the Appendix, as indicated in the sequel.
Here are our remarks:\vspace{0.1cm}
\begin{itemize}
\item[(A)] The VEV of the energy
density computed in \cite{Z} is affected by an error of sign, both in the case of two
impurities and in the case of a single impurity.
In fact, in  Eqs. (92a) (92b) of \cite{Z} laying the basis of such computations,
the right-hand sides should be replaced by their opposites (see subsection A
in the Appendix of this note). Due to this, the overall signs should be changed
in all the expressions for the energy density derived in \cite{Z}
after Eqs. (92a) (92b).\vspace{0.1cm}
\item[(B)] In the special case of a single
point-like impurity, described by Eq. (106) of
\cite{Z}, there is another minor error: besides changing the overall sign,
the factor $1/2$ in front of the integral giving the energy density VEV should
be replaced by $1/4$ (see subsection B in the Appendix of this note).
The energy density VEV at a space point $\vec{x} \in \mathbf{R}^3$ is denoted in Eq. (106)
of \cite{Z} with $\mathcal{E}_{single}(\vec{x},0)$, where
the subscript indicates the presence of a single impurity
and $0$ is the limit value of the scaling parameter in the
point-like setting. The corrected formulation of Eq. (106) in \cite{Z},
with the amendments indicated above, reads
\begin{equation}
\mathcal{E}_{single}(\vec{x}, 0) = {1 \over 4 |\vec{x}|^4} \int_{0}^{+\infty}\!\! d r\; {1 + 2 |\vec{x}| r \over \alpha + 2 \pi^2 r}\; e^{- 2 |\vec{x}| r} \,.
\label{eq:corrE}
\end{equation}
Here $\alpha >0$ is a parameter appearing in the
mathematical description of the impurity (this is proportional
via a purely numerical factor to the homonymous parameter employed in \cite{A},
see Eq. \eqref{eq:alA} in the Appendix).
\vfill\eject\noindent
{~}
\vskip 0.cm \noindent
\item[(C)] \textsl{The correct version of Eq. (106) of \cite{Z}, just given
in Eq. \eqref{eq:corrE} of item (B), matches exactly Eq. (60) of our paper \cite{FP}
for energy density VEV in the
case of a single point-like impurity (with conformal parameter $\xi=0$ for
the quantum field theory); this disproves Ziemian's claim (I)}.\par \noindent
Let us justify our statement. Firstly, let us recall
that in \cite{FP} we computed (for arbitrary values of the conformal
parameter $\xi$) the VEVs $\langle 0| \hat{T}_{\mu \nu} |0 \rangle_{ren}$,
renormalized by the zeta method \cite{FPB},
where $\hat{T}_{\mu \nu}$ ($\mu,\nu = 0,1,2,3$) are the
components of the stress-energy tensor (these are operators, as usual in second quantization).
The choice $(\mu, \nu) = (0,0)$ corresponds to the energy density.
According to Eq. (60) of \cite{FP} with $\xi=0$ (and to Eq. (58) of the same reference),
at any space point $\vec{x} \in \mathbf{R}^3$ we have
\begin{equation}
\langle 0| \hat{T}_{0 0} |0 \rangle_{ren}(\vec{x}) =  {1  \over 8 \pi^2 |\vec{x}|^4} \Big[ 1 + (1 - \rho)\, \mathfrak{E}(\rho)  \Big]_{\displaystyle{\rho = {2 |\vec{x}|/\gamma}}}\; , \qquad
\mathfrak{E}(\rho) := \int_{0}^{+\infty}\!\! d v \; {e^{- \rho v} \over 1 + v} \; .
\label{eq:T00ren}
\end{equation}
In the above, $\gamma >0$ is a parameter related to the formal
description of the impurity, with a role similar to that of the parameter
$\alpha$ in Ziemian's energy density $\mathcal{E}_{single}(\vec{x}, 0)$.
(In Ref. \cite{FP}, $|\vec{x}|$, $\gamma$ and $\mathfrak{E}$ were respectively
indicated with the symbols $r$, $\lambda$ and $\mathcal{E}$, here avoided
to prevent confusion with Ziemian's work \cite{Z} where $r$ is the typical name
of an integration variable, $\lambda$ is the scaling parameter and
$\mathcal{E}$ is used to denote the energy density).\par \noindent
Let us now reconsider Ziemian's energy density VEV $\mathcal{E}_{single}(\vec{x}, 0)$,
taking into account the corrections indicated before in item (B) (see Eq. \eqref{eq:corrE}
of this note).
To make a comparison with our result for the energy density VEV, it is convenient to make the change
of variable $\displaystyle{ r = {\alpha \over 2 \pi^2}\,v}$ in the integral giving
$\mathcal{E}_{single}(\vec{x}, 0)$; in this way, we obtain
\begin{equation}
\mathcal{E}_{single}(\vec{x}, 0) = {1  \over 8 \pi^2 |\vec{x}|^4}
\left[\, \int_{0}^{+\infty}\!\! d v\; {1 + \varrho v \over 1 + v}\; e^{- \varrho v} \right]_{\displaystyle{\varrho = \alpha |\vec{x}|/\pi^2}} \, .
\end{equation}
On the other hand, using the trivial identity $\displaystyle{1 + \varrho v \over 1 + v } = \varrho + \displaystyle{1 - \varrho  \over 1 + v }$ we infer
\begin{equation}
\int_{0}^{+\infty}\!\! d v\; {1 + \varrho v \over 1 + v}\; e^{- \varrho v}
= \varrho \int_{0}^{+\infty}\!\! d v\; e^{- \varrho v}
+ (1-\varrho) \int_{0}^{+\infty}\!\! d v\; {e^{- \varrho v}  \over 1 + v}
= 1 + (1 - \varrho)\, \mathfrak{E}(\varrho)\, ,
\end{equation}
with $\mathfrak{E}$ as in Eq. \eqref{eq:T00ren} of the present note. Thus
\begin{equation}
\mathcal{E}_{single}(\vec{x},0) =  {1  \over 8 \pi^2 |\vec{x}|^4}
\Big[  1 + (1 - \varrho) \mathfrak{E}(\varrho)  \Big]_{\displaystyle{\varrho =} \displaystyle{\alpha |\vec{x}|/\pi^2}} \, .
\end{equation}
Comparing this with Eq. \eqref{eq:T00ren}
for $\langle 0| \hat{T}_{0 0} |0 \rangle_{ren}$, we see that
\begin{equation}
\mathcal{E}_{single}(\vec{x},0) = \langle 0| \hat{T}_{0 0} |0 \rangle_{ren}(\vec{x})
\qquad \mbox{if\; $\alpha = \displaystyle{2 \pi^2 \over \gamma}$}\;.
\label{eq:EEeqT00}
\end{equation}
Let us remark that the above relation between the parameter $\alpha$
of Ziemian and our parameter $\gamma$ can be derived as well by an independent argument,
reported in subsection C of the Appendix to this note.

\item[(D)] Let us pass to paper \cite{F},
where one of us (D.F.) discussed the total energy VEV
in the case of a single point-like impurity.
We believe that Ziemian's description of \cite{F}
(see item (II) in the Introduction) is oversimplified
and requires a clarification, even in view of the
connections between \cite{F} and the literature cited therein.
As a matter of fact, Ziemian
overlooks the decomposition of the total energy as
the sum of a \textsl{bulk} and of a \textsl{boundary} energy, given
in \cite{F} following general prescriptions from  \cite{FPB}.
Upon renormalization via zeta regularization (including subtraction
from the energy density of its empty space analogue), a finite value is obtained
in \cite{F} for the bulk energy; as indicated therein, this
result agrees with a previous computation by Spreafico and Zerbini \cite{S}
(in the terminology of \cite{F}, these authors did not consider the boundary
energy and approached directly the bulk energy via a global
zeta method). Admittedly,
the boundary contribution to the total energy individuated
in \cite{F} remains divergent even after renormalization. Besides
the considerations proposed therein on this fact, it can
be useful to mention that the persistent
singular behavior of certain renormalized observables,
related to the presence of classical boundaries,
is a well known pathology of quantum field
theories. To some extent, the model with a delta impurity
can be regarded as an example of a quantum
field theory on a domain with classical
boundaries, since the mathematical description
of delta-type potentials is actually attained by means
of suitable boundary conditions.
\end{itemize}
\section*{Appendix}
Let us review some notations of \cite{Z} that will appear frequently in
the present Appendix, and which were partly mentioned in Section \ref{disc}
of this note. \par \noindent
The non-local pseudo-potential modeling the impurities in \cite{Z}
depends on: a vector $\vec{a} \in {\bf R}^3$
determining the centers of the impurities,
which are placed at points $\pm \, \vec{a}/2$, or at points
$\vec{0}$ and $\vec{a} \to \infty$ to recover
in this limit the case with a single impurity; a shape function $g$; a parameter $\alpha >0$
(see Sections 2 and 3 therein). Section 4 of \cite{Z} considers
a rescaled model governed by an additional parameter $\lambda >0$; in the
limit $\lambda \to 0^{+}$ the impurities become point-like,
and one recovers a local potential consisting of two
Dirac deltas
or just one, if the second impurity is sent to infinity. \par \noindent
We already indicated that the role of $\alpha$ in \cite{Z}
is similar to that of the homonymous parameter in the book
of Albeverio \textsl{et al.} \cite{A}; in the present Appendix,
to avoid confusion we indicate with $\alpha_A$ the parameter
of \cite{A}. For future use, we mention that a comparison between Sections 2-4 of
\cite{Z} and Chapter II.1 of \cite{A} gives
\begin{equation}
\alpha = 8 \pi^3 \alpha_A \,. \label{eq:alA}
\end{equation}
The forthcoming subsections A,B,C provide support and/or additional
information for our statements (A)(B)(C) in Section \ref{disc} of this note.
\vspace{-0.15cm}
\subsection*{A. On Eqs. (92a) and (92b) of \cite{Z}}
In Section \ref{disc} of this note, item (A) claims that the cited Eqs. (92a) and (92b) have wrong signs,
meaning that their right-hand sides should be replaced by
their opposites; let us justify this statement. Eqs. (92a)(92b) are presented in \cite{Z}
as operator analogues (arising from spectral resolution
methods) of certain elementary identities about real numbers, described
in Eq. (91). In building such operator analogues
one should essentially replace the numbers $(r^2 + k^2)^{-1}$ or
$(r^2 + p^2)^{-1}$ of Eq. (91) ($r,k,p > 0$)
with the inverse operators $(r^2 + h^2_{\vec{a}})^{-1}$
or $(r^2 + h^2)^{-1}$, where $r$ is again a positive
real number and $h_{\vec{a}}, h$ are the basic Hilbert space operators
employed throughout \cite{Z}.
The author of \cite{Z} seems to identify the operators $(r^2 + h^2_{\vec{a}})^{-1}$
and $(r^2 + h^2)^{-1}$ with
the operators $G(-r^2)$ and $G_0(-r^2)$,
defined following Eq. (17) of that work. Unfortunately,
Eq. (17) of \cite{Z} gives
$G(-r^2) = (-r^2 - h^2_{\vec{a}})^{-1} = - (r^2 + h^2_{\vec{a}})^{-1}$
and, similarly, $G_0(-r^2) = - (r^2 + h^2)^{-1}$.
({\footnote{Let us make a further comment. Apart from the sign problem,
it is not so evident that Eq. (92a) for $h_{\vec{a}} -h$ can be obtained via spectral
methods as an operator analogue of the second identity in Eq. (91), since
$h_{\vec{a}}$ and $h$ do not commute.
However, we have found the following alternative derivation of Eq. (92a)
with the correct sign:
for any given $k,p>0$, instead of using the second
identity in Eq. (91) for $k-p$, write the separate identities
$k = - (2/\pi) \int_{0}^{+\infty}\! d r\, \big[r^2 (r^2 + k^2)^{-1} - 1\big]$,
$p = - (2/\pi) \int_{0}^{+\infty}\! d r\, \big[r^2 (r^2 + p^2)^{-1} - 1\big]$,
consider their operator analogues with $k,p$ replaced
by $h_{\vec{a}}, h$, and finally take their difference.\vspace{-0.2cm}}})
\par \noindent
We already indicated in Section \ref{disc} of this
note that, as a consequence of the error of sign
in Eqs. (92a) (92b), all expressions for
the energy density VEV given in \cite{Z} after  Eqs. (92a) and (92b)
should be replaced by their opposites.
\vfill\eject\noindent
{~}
\vskip 0.3cm \noindent
\subsection*{B. On Eq. (106) of \cite{Z} for \boldmath{$\mathcal{E}_{single}(\vec{x},0)$}}
The cited equation gives the VEV of the energy density
for a single, point-like impurity.
Item (B) in Section \ref{disc} of this note claims that,
besides the wrong overall sign (see the end of subsection A),
this equation has a wrong factor $1/2$ before the
integral appearing therein, to be replaced with $1/4$.
Let us check this statement, following carefully the setting of
\cite{Z} for such computations. \par \noindent
The energy density VEV of Eq. (106) at a space point $\vec{x} \in \mathbf{R}^3$
is the limit
\begin{equation}
\mathcal{E}_{single}(\vec{x}, 0) := \lim_{\lambda \to 0^+} \mathcal{E}_{single}(\vec{x}, \lambda)\,.
\end{equation}
Here,  $\mathcal{E}_{single}(\vec{x}, \lambda)$ is
the energy density VEV for the
rescaled model with a single, extended impurity; $\lambda >0$ is the scaling parameter already mentioned
before, and we recall that the limit $\lambda \to 0^{+}$
corresponds to a point-like impurity. \par \noindent
According
to Eq. (99) of \cite{Z} (with a change of sign, see again subsection A in this Appendix) and to the instructions of
\cite{Z} on 
the rescaling procedure (see the first lines of subsection 8.1 therein), one has
\begin{equation}
\mathcal{E}_{single}(\vec{x}, \lambda) = {1  \over 2 \pi} \int_{0}^{+\infty} \!\!d r\;
{H'_{\lambda}(r, |\vec{x}|)^2 - r^2 H_{\lambda}(r,|\vec{x}|)^2 \over t_{\lambda}(i r)}\; ,\label{ret} 
\end{equation}
where $H_{\lambda}$, $t_{\lambda}$ are certain functions,
and $H'_{\lambda}$ is the derivative of $H_{\lambda}$ w.r.t
the second variable. Eq. (105) of \cite{Z} states that
\begin{equation}
H_{\lambda}(r,|\vec{x}|) = \sqrt{{\pi \over 2} } \; \hat{g}(i \lambda r)\; {e^{- |\vec{x}| r} \over | \vec{x} |} \; , \quad
H'_{\lambda}(r,|\vec{x}|) = - \,\sqrt{{\pi \over 2} } \; \hat{g}(i \lambda r)\; {(1 + |\vec{x} | r)\, e^{- |\vec{x}| r} \over | \vec{x} |^2}\;,
\end{equation}
where $\hat{g}$ is, essentially, the Fourier transform of the
shape function $g$ mentioned in the first lines of this Appendix.
Moreover, Eq. (11) and Eq. (35a) of \cite{Z} (with $w = i r$) give
\begin{equation}
t_{\lambda}(i r) = \alpha + 2 \pi \int_{-\infty}^{+\infty}\!\! d p \; {r^2 \ \over r^2 + p^2} \; \hat{g}(\lambda p)^2 \,.
\end{equation}
Let us consider the $\lambda \to 0^{+}$ limits of the above functions,
indicated replacing the subscript $\lambda$ with $0$. Using the fact
that $\hat{g}(0) = 1$ (see Eq. (14) of \cite{Z}),
we get
\begin{equation}
\displaystyle{ H_{0}(r,|\vec{x}|) = \sqrt{{\pi \over 2} } \; {e^{- |\vec{x}| r} \over | \vec{x} |} \;,
\quad
H'_{0}(r,|\vec{x}|) = -\, \sqrt{{\pi \over 2} } \; {(1 + | \vec{x} | r)\, e^{- |\vec{x}| r} \over | \vec{x} |^2}\; , } 
\end{equation}
\begin{equation} 
\displaystyle{ t_{0}(i r) = \alpha + 2 \pi \int_{-\infty}^{+\infty}\!\! d p\; {r^2 \over r^2 + p^2} = \alpha + 2 \pi^2 r \; .}
\end{equation}
So, returning to the expression of $\mathcal{E}_{single}(\vec{x},\lambda)$ (Eq. \eqref{ret} in this Appendix) and
considering the limit $\lambda \to 0^{+}$, we obtain
\begin{align*}
\mathcal{E}_{single}(\vec{x}, 0)
& = {1 \over 2 \pi}\! \int_{0}^{+ \infty}\!\! d r\,
{1 \over \alpha + 2 \pi^2 r } \left[
\left( -\,\sqrt{\pi \over 2}\, {(1 + | \vec{x} | r)\, e^{- |\vec{x}| r} \over | \vec{x} |^2} \right)^{\!2}
- r^2 \left( \sqrt{\pi \over 2 }\, {e^{- |\vec{x}| r} \over |\vec{x} |} \right)^{\!2}
\right] \\
& = {1 \over 4 |\vec{x} |^4}\! \int_{0}^{+\infty}\!\! d r\; {1 + 2 |\vec{x}| r \over \alpha + 2 \pi^2 r}\; e^{- 2 |\vec{x}| r} \,.
\end{align*}
This is the correct formulation of Eq. (106) for $\mathcal{E}_{single}(\vec{x}, 0)$, also reported
in Eq. \eqref{eq:corrE} of this note.
\vfill\eject\noindent
{~}
\vskip 0.2cm \noindent
\subsection*{C. On the relation \boldmath{$\alpha = 2 \pi^2/\gamma$}}
This relation appears in Eq. \eqref{eq:EEeqT00}
of this note, where it is derived a posteriori
as the condition for equality between the (corrected) energy density
VEV $\mathcal{E}_{single}(\vec{x},0)$ of Ziemian's paper \cite{Z} and
the energy density VEV  $\langle 0| \hat{T}_{0 0} |0 \rangle_{ren}(\vec{x})$
of our paper \cite{FP} (with vanishing conformal parameter). \par \noindent
As a matter of fact, the above relation between $\alpha$ and $\gamma$
can be established a priori, by an argument unrelated
to the comparison of the energy densities of \cite{Z} and
\cite{FP}. To this purpose, recall that the parameter $\alpha$
of \cite{Z} and the homonymous parameter in \cite{A}, here indicated with $\alpha_A$
to avoid confusion, are related by $\alpha = 8 \pi^3 \alpha_A$ (see Eq. \eqref{eq:alA}
at the beginning of this Appendix).
On the other hand, again by comparison with \cite{A}, our paper \cite{FP} specifies
that $\alpha_A = 1/(4 \pi \gamma)$ (see the comment in the fifth line
after Eq. (23) of \cite{FP}, recalling that $\gamma$ is indicated
therein with $\lambda$). Summing up we have $\alpha = 2 \pi^2/\gamma$,
in agreement with Eq. \eqref{eq:EEeqT00} of this note.
\section*{Acknowledgments}
This work was supported by: INdAM, Gruppo Nazionale per la Fisica Matematica;
INFN; MIUR, PRIN 2010 Research Project “Geometric and analytic theory of Hamiltonian
systems in finite and infinite dimensions.”; Universit\`{a} degli Studi di Milano;
Progetto Giovani INdAM-GNFM 2020 “Emergent Features in
Quantum Bosonic Theories and Semiclassical Analysis”.

\vfill \eject \noindent
\end{document}